\documentstyle[11pt,aaspp4]{article}

\tighten
\eqsecnum

\accepted{29 June 1998}

\slugcomment{Accepted in Astronomical Journal.}

\begin{document}

\title{Bulge-Disk Decomposition of 659 Spiral and Lenticular\\
       Galaxy Brightness Profiles}

\author{W. E. Baggett}
\affil{Science Programs, Computer Sciences Corporation \\ 3700 San Martin Drive, 
       Baltimore, MD 21218\\  baggett@stsci.edu}
\author{S. M. Baggett}
\affil{Space Telescope Science Institute \\ 3700 San Martin Drive, 
       Baltimore, MD 21218 \\ sbaggett@stsci.edu}
\author{K. S. J. Anderson}
\affil{Department of Astronomy, New Mexico State University, \\ P. O. Box 30001,
       Dept. 4500, Las Cruces, NM 88003 \\ kurt@nmsu.edu }




\begin{abstract}

We present one of the largest homogeneous sets of spiral and
lenticular galaxy brightness profile decompositions completed to date.
The 659 galaxies in our sample have been fitted with a deVaucouleurs'
law for the bulge component and an inner-truncated exponential for the
disk component. Of the 659 galaxies in the sample, 620 were
successfully fit with the chosen fitting functions.  The fits are
generally well-defined, with more than 90\% having RMS deviations from
the observed profile of less than 0.35 magnitudes. We find no
correlations of fitting quality, as measured by these RMS residuals,
with either morphological type or inclination. Similarly, the
estimated errors of the fitted coefficients show no significant trends
with type or inclination.

These decompositions form a useful basis for the study of the light
distributions of spiral and lenticular galaxies.  The object base is
sufficiently large that well-defined samples of galaxies can be
selected from it.

\end{abstract}


%
%

\section{Introduction}

In order to compare the large-scale characteristics of galaxies
objectively, quantitative measures of the structural components are
necessary. There are many schemes for describing the structure of
galaxies, including Hubble classification (Sandage\markcite{san61},
1961), isophotal radii (Holmberg\markcite{hol47}, 1947), concentration
parameters (Kent\markcite{ken85}, 1985; Kodaira et
al.\markcite{kod90}, 1990), and the use of various fitting functions
(deVaucouleurs\markcite{dev53}, 1953; Freeman\markcite{fre70}, 1970;
Kormendy\markcite{kor77}, 1977). All these techniques specify
parameters which can provide insight into the formation and evolution
of galaxies.  The use of standardized fitting functions is arguably
the most powerful method for measuring the large-scale structure of
galaxies, as the functions yield a variety of parameters which can be
easily compared with the results of theoretical models.  They also
provide a reasonably detailed description of the radial light
distribution with a small number of parameters.

Ideally, fitting functions would be based upon the physics of the
formation and evolutionary processes. Unfortunately, these processes
are neither simple nor well-understood, so the most commonly used
functions are derived empirically. Traditional fitting functions for
elliptical galaxies and spiral galaxy bulges include the Hubble law
(Hubble\markcite{hub30}, 1930), King model (King\markcite{66}, 1966),
and deVaucouleurs law (deVaucouleurs\markcite{dev53}, 1953).
Recently, there has been some work which suggests that a generalized
version of the deVaucouleurs profile ($r^{1/n}$) provides for better
bulge fits (Andredakis, Peletier, and Balcells\markcite{and95}, 1995),
and that late-type spirals often have bulges which are best fitted by
exponentials (Andredakis and Sanders\markcite{and94}, 1994).
Exponentials (Freeman\markcite{fre70}, 1970) and inner-truncated
exponentials (Kormendy\markcite{kor77}, 1977) work well for the disk
components of spiral galaxies. Overall, the deVaucouleurs law seems to
be quite effective as a fitting function for bulges; it can be written
as

\begin{equation}
I_{\rm B}(r) = I_{\rm e} \times 10^{-3.33((r/r_{\rm e})^{1/4} - 1)}
\end{equation}

\noindent 
where $I_{\rm B}(r)$ is the surface intensity of the bulgeat radius
$r$, $r_{\rm e}$ is a characteristic radius defined to be the radius
within which half the total light is emitted, and $I_{\rm e}$, the
effective intensity, is simply the surface intensity at $r_{\rm e}$.
(In this paper, we will consistently use the term "surface intensity,"
$I$, to mean linear intensity units per square arcsecond, and "surface
brightness," $\mu$, to mean the same quantity in magnitude units.)
Similarly, the inner-truncated exponential is defined by
Kormendy\markcite{kor77} (1977) as

\begin{equation}
I_{\rm D}(r) = I_{\rm 0} \times \exp{-(r/r_{\rm 0} + (r_{\rm h}/r)^n)}
\end{equation}

\noindent
where $I_{\rm D}(r)$ is the disk surface intensity at radius $r$,
$I_{\rm 0}$ is the central intensity of the disk, $r_{\rm 0}$ is the
disk scale length, and $r_{\rm h}$ is the radius of the central cutoff
of the disk ("hole radius"); the pure exponential disk is the same as
equation (1-2) with $r_{\rm h}=0$.  Kormendy\markcite{kor77} (1977)
found that a value of $n\sim3$ in the truncation term works well, and
we have adopted $n=3$ for all of the fits presented here.  Figure 1
illustrates the usefulness of including a truncation term in the
fitting function, using NGC~3145 as an example.

Others have used these fitting functions to obtain the relevant
structural parameters for spiral galaxies in order to compare galaxies
of different types, luminosities, and environments. For example,
Boroson\markcite{bor81} (1981) fit brightness profiles for 26
non-barred spiral galaxies in order to determine how the bulge-to-disk
ratios are related to the Hubble types, and to investigate the
relationship between spiral and S0 (lenticular) galaxies.
Kent\markcite{ken85} (1985) performed a similar analysis using 105
intrinsically luminous galaxies of all types. Kormendy\markcite{kor77}
(1977) decomposed the brightness profiles of seven compact S0 galaxies
and one "normal" galaxy to check Freeman's\markcite{fre70} (1970)
hypothesis of a constant central disk surface brightness.  More
recently, deJong\markcite{dej96} (1996) investigated the
Freeman\markcite{fre70} (1970) result of a constant central surface
brightness of disks, and some other relationships between the fitting
parameters and the Hubble sequence, using $B-$ and $K-band$ brightness
profiles of 86 face-on disk galaxies.

A relatively recent innovation is to fit a surface to the entire
galaxy image (Byun and Freeman\markcite{byu95} 1995) using the above
fitting functions and also solving for the center and ellipticity of
the projected distributions.  A general advantage of this approach is
that the bulge and disk components can be allowed to have different
ellipticities, which alleviates the problem of projection effects for
moderate- to high-inclination systems: because the rounder bulge is
typically sampled at a smaller galactocentric radius than the inclined
disk for a given position in the image, the derived bulge parameters
are systematically too large when estimated from brightness profiles
obtained by azimuthal averaging techniques.  However, the profiles
used here are major axis cuts (see below), so this difficulty should
not affect our fitted parameters (Burstein\markcite{bur79} 1979). The
cost of using major axis cuts is that of throwing away much of the
information in the images.

All of these programs except Kormendy\markcite{kor77} (1977) employed
a simple exponential to describe the disk light distribution. As part
of a study of the origin of inner-truncated spiral galaxy disks, or
Type II brightness profiles (Freeman\markcite{fre70}, 1970), we have
used the deVaucouleurs' law and the inner-truncated disk fitting
function from Kormendy\markcite{kor77} (1977) for the bulge-disk
decomposition of 659 spiral and lenticular galaxy brightness
profiles. Our preliminary study (Baggett, Baggett, and
Anderson\markcite{bag93}, 1993) indicated that a substantial fraction
of all spiral galaxies exhibit an inner-truncation, so the inclusion
of such a term in the fitting function seems justified with this large
set of brightness profiles.  Furthermore, the data set used in this
study is extremely homogeneous, all images having been obtained,
reduced, and analyzed in the same way.  Thus, the results of our
fitting should be a useful resource for many studies of the
large-scale properties of disk galaxies. The following sections will
describe the data and the bulge-disk decomposition procedures, and
will present the fitting results together with a discussion of the
associated errors.

\section{Data}

The brightness profiles used for this study were obtained from Kodaira
et al.\markcite{kod90} (1990; hereafter, PANBG) in a machine-readable
form.  The initial PANBG sample of galaxies was selected on the basis
of being included in {\em A Revised Shapley-Ames Catalog of Bright
Galaxies} (Sandage and Tammann\markcite{san81}, 1981; hereafter, RSA)
and being north of declination -25\arcdeg. Of the 911 such galaxies in
the RSA, 791 are included in the PANBG, and 659 of those have Hubble
types (T-Types) from the {\em Third Reference Catalogue of Bright
Galaxies} (deVaucouleurs et al.\markcite{dev91}, 1991; hereafter RC3)
in the range -3 to 9, which indicates that they are spiral or
lenticular galaxies. These 659 galaxies form the basis of our study.

Galaxies in the PANBG were observed photographically over a period of
almost two decades (late 1970's through 1988) with the Kiso
Observatory 1.05-m Schmidt telescope, using Kodak IIa-D plates and a
Schott GG495 glass filter to define the "photographic V-band."
Exposure times ranged from 30 minutes to 60 minutes, with 50 minutes
being standard, and the plates were developed in Fuji Pandol or Kodak
D-19 developer. The plates were then digitized with the Kiso
Observatory PDS microdensitometer utilizing a 1\arcsec\ square
aperture, except for NGC~224 and NGC~598, which were measured with a
10\arcsec\ square aperture. Each plate included a step wedge image
which was scanned in the same fashion as the galaxy images. Measured
densities were converted into relative intensities via the wedge
calibrator, and the aperture photometry from Longo and
deVaucouleurs\markcite{lon83} (1983) was then used to transform the
resulting magnitudes to a standard photometric system. The stated
internal photometric accuracy in the PANBG is about 0.1 magnitudes
(standard deviation), and is dominated by errors in the absolute
calibration.

The brightness profiles in the PANBG were obtained from the resulting
calibrated images by taking a cut along the apparent major axis of
each galaxy.  The major axis was defined by fitting the 25
$V-mag~arcsec^{-2}$ isophote with an ellipse whose center was fixed at
the center of gravity of a 21x21 pixel region around the apparent
nucleus of the galaxy. The surface brightness was then sampled along
this axis using a circular aperture which was stepped outward from the
galaxy center in 2-pixel steps (2\arcsec\ for all but NGC~224 and
NGC~598, which used 20\arcsec\ steps). The aperture used was 2 pixels
in diameter at the galaxy center and the diameter was varied in such a
way as to be tangent to a sector with a 5 degree vertex angle centered
on the major axis.  This scheme of varying the aperture size was
chosen in order to compensate for the decreased signal-to-noise ratio
in the outer portions of the galaxies. As a result, there is some
radial smearing of the intensity information at large galactocentric
radii, smoothing structure in the outer portions of each profile.
Further smoothing results from our averaging of the two halves of the
major axis cut to produce the final profiles.  For full details of the
data acquisition and reduction processes, the reader is referred to
PANBG.

\section{Fitting}

\subsection{Procedure}

The major axis brightness profiles were fitted using a combination of
a deVaucouleurs'\markcite{dev53} (1953) law (equation 1-1) and an
inner-truncated exponential (Kormendy\markcite{kor77}, 1977; equation
1-2).  The interactive STSDAS task 'nfit1d' was used for all of the
fitting; this task uses the downhill simplex algorithm for performing
a non-linear least-squares fit of the data to a specified function,
and allows interactive control over the inclusion of the various
parameters and the range of the data to be fit. It is a very flexible
routine and we found that it accurately returns the values of the
fitting parameters in a number of test cases. Fitting is performed on
the surface intensity data and is accomplished by minimizing the
weighted RMS deviation of the data from the fit.

The most appropriate weighting function, $w_{i}$, is one which uses
the variance of the intensity measurement for each point as its basis,
with the weight of the $i^{th}$ point being

\begin{equation}
w_{i} = \frac{1}{\sigma_{i}^{2}}
\end{equation}

\noindent 
where $\sigma_{i}^{2}$ is the variance of the $i^{th}$ point
(Bevington\markcite{bev69}, 1969). We chose to use a weighting based
on the Poisson distribution, where $\sigma_{i}\propto\sqrt{I}$, as
this was consistent with the fashion in which the intensity
measurements were obtained.

An unfortunate side effect of this $1/I_{i}$ weighting function is
that it destroys the usefulness of the weighted RMS residual as a
goodness of fit measure {\em between} galaxies. The value of the
weighted RMS residual is highly dependent upon the fitting range, with
fits to lower surface intensities being virtually guaranteed a lower
weighted RMS value than fits stopping at higher surface
intensities. As a result, the weighted RMS residual is a useful
diagnostic only during the fitting process for a given galaxy and, as
such, the weighted RMS values for each fit are not reported here.

However, the {\em unweighted} RMS residuals, $\sqrt{[\sum(\mu_{i} -
\mu_{fit})^{2}]/N}$, can be computed after the fact, and we have
tabulated these values, expressed in magnitude units, as a basis for
assessing the relative quality of the fits.  These residuals were
computed from the portions of each brightness profile at radii larger
than 3\arcsec ~and out to the point where the profile first drops to a
surface brightness fainter than 25 $V-mag~arcsec^{-2}$. Hence, all of
the calculations avoid the portion of the profile most affected by
seeing, and reach the same limiting surface brightness. Also, any
structure present in the profile contributes to this measure, and thus
galaxies with significant structure will be recognizable by their
correspondingly larger RMS value. In this way, the unweighted RMS
residuals are directly comparable from galaxy to galaxy, and reflect
more accurately the quality of the fits than do the weighted RMS
residuals.

Aside from the weighting function, the only other controlled aspect of
the fitting was the choice of components to include in each fit.
Generally, all fits were attempted with both a bulge and a pure
exponential disk, resulting in the estimation of four quantities and
their uncertainties: $I_e$, $r_e$, $I_0$, and $r_0$; inner-truncated disks,
with the additional parameter, $r_h$, were only utilized if the
profile had the suggestion of a plateau near the center.  If any
component (bulge, disk, inner-truncated disk) was fit with a negative
value for a coefficient, that component was deemed non-physical and
was removed from the fitting function. In cases where the need for a
specific component was not obvious, fits with and without it were
obtained and the component was included if the {\em weighted} RMS
value was smaller by at least 10\% than the fit without the component.

A sample of the results of the fitting are presented in Table 1, which
provides for each galaxy its NGC/IC designation (as indicated by a
leading "N" or "I", respectively), its revised Hubble type, T-type,
and axis ratio from the RC3, and the fit parameters determined
here. The fit parameters are given as the fitting range in arcseconds
(in the format minimum:maximum), the bulge effective surface
brightness ($\mu_{\rm e}$), the bulge effective radius ($r_{\rm e}$),
the disk central surface brightness ($\mu_{\rm 0}$), the disk scale
length ($r_{\rm 0}$), and the disk hole radius ($r_{\rm h}$).  The
table also includes the seeing as reported in the PANBG, and the
seeing-corrected values of the fitted bulge parameters ($\mu^{0}_{\rm
e}$ and $r^{0}_{\rm e}$; see the Appendix). The last two columns
contain the unweighted RMS deviation of the fit from the profile in
$V-mag~arcsec^{-2}$ and a column of notes.  All surface brightness
quantities are in units of $V-mag~arcsec^{-2}$, and all radii are in
units of arcseconds.  No corrections for galactic extinction, internal
extinction, or inclination have been applied --- the fits are for the
observed major axis profile.  We have chosen to present the results
for the observed profiles in order to allow the reader the opportunity
to apply the corrections (s)he deems most appropriate; we therefore
avoid the necessity of uncorrecting our fits and subsequently applying
a different correction.

The fitting procedure did not include any allowance for the effects of
seeing on the brightness profile.  In the Appendix, we describe an
experiment which was designed to quantify the impact on the fitted
bulge parameters of excluding seeing from the fitting; the net result
is that our fitted $I_{e}$ tends to be too large by up to an order of
magnitude when the seeing is much larger than the input effective
radius, and our fitted $r_{e}$ tend to be slightly too small in the
same cases.  This is what is intuitively expected, of course.

Although the profile definition procedure introduces a significant
level of radial smearing, structure is still apparent in many of the
profiles. The fitting process nominally made no allowance for the
presence of structure, fitting across arms and bars as if they were
simply noise in the data, except in some specific instances. These
instances occur when a very strong, isolated feature is present ---
then the radial range occupied by the feature was excluded from the
fit. If a strong feature is present at the end of the fitting range,
the fitting range was suitably shortened.  The existence of such a
condition was manually determined, and is indicated in Table 1 by the
presence of multiple ranges in the fitting range column and/or by a
note ("A") in the last column of the table.

Figure 2 presents the profiles and fits for the galaxies from Table 1.
Each plot shows the observed profile from the PANBG as crosses, and
the fitted bulge and disk components and their sum as solid lines.  In
addition, the range of radius included in the fit is indicated by the
horizontal line(s) at the 25.3 $V-mag~arcsec^{-2}$ level, and the fit
parameters are provided near the top of each plot.  This selection of
objects includes some very good fits as measured by the RMS deviations
(NGC~0016, NGC~0404), some typical-quality fits (NGC~0224, NGC~0237),
some fits which avoid strong structure in the profile (NGC~0151,
NGC~0157, NGC~0253), and one of the worst fits in the entire sample
(NGC~0289).

There were 39 galaxies in the sample for which no fits were obtained;
these objects are listed in Table 2, with their revised types and axis
ratios from the RC3.  These galaxies are particularly ill-suited for
fitting by the selected fitting functions, as many of them have
pronounced concavities toward low surface brightness or multiple
exponential components in their brightness profiles -- the chosen
functions simply do not represent their light distributions in any
meaningful way.

Finally, there are also several fits (e.g., NGC~0016, NGC~0628,
NGC~0890, and NGC~5033) where the disk component is everywhere fainter
than the bulge, and other fits where the disk becomes brighter than
the bulge only at intermediate radii (e.g., NGC~0670, NGC~0955,
NGC~1090, and NGC~1187).  In both of these situations, the bulge fit
might have benefitted from an alternative functional form, perhaps the
generalized deVaucouleurs' law as discussed by Andredakis, Peletier,
and Balcells\markcite{and95} (1995), although no attempt has been made
to investigate this in the present study.

\subsection{Fitting Errors}

We have chosen three methods of estimating the fitting errors: 1)
unweighted RMS residuals, 2) error estimates provided by the fitting
software, and 3) comparison with the results of other workers.

The unweighted RMS residuals for each fit are tabulated in Table 1.
Because we have computed them in a uniform fashion for all objects,
including all structure in the profiles, these values provide an
unbiased and consistent measurement of how well the fitting functions
and the determined parameters describe the observed brightness
distribution.  From these values we find that the median RMS deviation
of the fits is only 0.15~mag, and that more than 90\% of the fits are
better than 0.35~mag; Figure 3 shows the distribution of the RMS
deviations using bins of width 0.05~magnitude. The overall ability of
the fits to describe the brightness distributions is quite good, given
that the unweighted RMS residuals include all of the structure present
in the profile.

In an effort to quantify the value of the unweighted residuals as a
measure of the goodness of fit, we have examined plots of the profiles
and selected the best examples of profiles without significant
structure and which appear to be well represented by the chosen
fitting functions.  Note that this selection did not involve
consideration of the computed unweighted RMS residuals.  There were 16
objects included in this selection, and they have a mean unweighted
RMS deviation of 0.05~magnitudes and a range of 0.03~magnitudes to
0.10~magnitudes; this suggests that fits with
unweighted RMS residuals greater than about 0.10~magnitudes are
affected to some degree by significant structure and/or poor fitting
quality.  The full sample contains 161 galaxies with RMS
residuals of 0.10~magnitudes or less.

At the other end of the distribution, the worst-fit galaxies have been
investigated to attempt to find out why they were so poorly fit.  We
have inspected the 18 galaxies which have unweighted RMS residuals
greater than 0.5~magnitudes to search for common characteristics such
as morphology and inclination.  The T-Type distribution of these 18
galaxies is essentially the same as for the sample as a whole, so
there appears to be no correlation between poor fit and T-Type.  There
also appears to be no serious trend with inclination: the mean
inclination of the group is 57\arcdeg, consistent with a random
distribution of tilts.  The most common characteristic is a
low-surface brightness extension to the brightness profile, such that
the fit falls below the measurements in the outer portion. The
extension is sometimes featureless, sometimes it contains a distinct
bump (as if an outer ring or arm); sometimes it is nearly constant
brightness and other times it is more or less parallel with the inner
profile.  There is only one case (NGC~0157) where the problem region
is in the main portion of the disk, and this galaxy looks disturbed,
almost as if undergoing a collision. The apparent overriding reason
for the poor fits is simply that the chosen fitting functions do not
work well for some galaxies.  This same conclusion holds for the 39
galaxies which were not fitted in this effort: common characteristics
of these objects are extremely strong, large-scale structures,
concavity of the brightness profile toward faint surface brightness,
and apparent multiple components of the profile, usually with more
than one exponential.  This occasional inappropriateness of the
fitting functions suggests that careful consideration of the fitted
parameters and their error estimates should be exercised before
judging whether a specific fit is truly meaningful for a detailed
study of any individual galaxy.

The second method of judging the fitting errors was the use of
estimated coefficient errors as produced by the fitting software. The
'nfit1d' task estimates the errors in each parameter by a process of
bootstrap resampling, with a choice of distribution functions for use
in the procedure. We used the Poisson distribution as the distribution
function for the parameter error estimation as this reflects the
photon statistics expected to affect the measured relative intensities
in the brightness profiles.  In order to minimize the effects of
comparing parameters which vary wildly in value from galaxy to galaxy,
we have computed the fractional error in each fitted parameter. Table
3 lists these fractional error estimates for a sample of the galaxies
--- column 1 gives the galaxy identification and columns 2 through 6
the fractional error of each of the parameters.  Note that since the
fitting was performed in the surface intensity scale (not surface
brightness), the relative errors are computed in linear units, not
magnitude units.

The fractional errors for the full set of galaxies are summarized in
Table 4, which gives for each parameter the number of error estimates,
the mean fractional error, the standard deviation of the fractional
errors, the median fractional error, and the minimum and maximum
fractional errors.  As can be seen in Table 4, every parameter {\em
except} $I_0$ is significantly affected by outliers; histograms of the
fractional error of each parameter are shown in Figure 4, which
demonstrate this problem.  Removing the most discrepant outlier from
the satistical calculations (second part of Table 4) improves the
results considerably.  Unfortunately, there is not just one "bad"
fit in the sample causing all of these outlying points: the bulge
parameter errors are both dominated by the fit for NGC~2441, while the
disk central surface intensity error is affected by the fit for
NGC~2541, the disk scale length error by NGC~5033, and the hole radius
error by NGC~2997.

The bulge component of NGC~2441 dominates the disk only at radii
smaller than about 6 arcsec, so there are very few data points
defining the bulge and the coefficient uncertainties reflect this
fact.  The profile for NGC~2541 is fitted with a faint inner-truncated
disk (the peak disk brightness is only about $24~V-mag~arcsec^{-2}$)
and the hole radius is more than six disk scale lengths from the
galaxy center.  The southern half of the brightness profile contains a
relatively bright spiral arm (about $23~V-mag~arcsec^{-2}$ at its
brightest), and this asymmetry results in a bump in the averaged
profile which has been fitted with the inner-truncation.  NGC 5033 was
fitted with a very faint ($\mu_0 = 25.1~V-mag~arcsec^{-2}$), very flat
disk ($r_0$ = 732 arcsec) which was never comparable in brightness to
the bulge.  As a result, the disk parameters for this galaxy are not
well constrained by the data, and the estimated errors in the
coefficients are correspondingly large.  Finally, NGC~2997 has been
fitted with a hole radius just smaller than the radius of the
innermost data point point, so the value of the hole radius is, again,
not really constrained by the data.

Because of the presence of these extreme outliers, the median
fractional errors are much more useful than the mean for examining the
fitting errors of the sample as a whole.  The summary in Table 4
shows that the median coefficient fractional errors range from a low
of 1.6\% for the disk scale length to more than 12\% for the bulge
effective intensity. From these data, it is clear that the bulge
parameters are the least well determined quantities in the fits, while
the disk parameters are generally well determined.  This is not
unexpected: since the bulge coefficients are usually dominated by a
relatively small number of data points, the constraints on them are
not very strong.

Finally, a few of the galaxies included in this study have been
previously fitted by others with the same fitting functions, providing
a completely independent check of the results of our fitting
procedure. There are five galaxies in the current study which are also
included in Boroson\markcite{bor81} (1981), and 12 are in common with
Kent\markcite{ken85} (1985), making a total of 17 measurements
available for use in this comparison.  Prior to making any comparisons
between the various works, the surface brightness parameters reported
in Boroson\markcite{bor81} (1981) and Kent\markcite{ken85} (1985) have
been "uncorrected" for the effects of inclination and galactic
absorption as applied in each study, and the length parameters have
been converted from kiloparsecs to arcseconds using the distances
adopted by those authors.  Furthermore, Boroson\markcite{bor81} (1981)
lists values for the disk $B-V$ color, which have been used to convert
his disk central surface brightnesses from the B bandpass to the V.
Finally, there are two galaxies in common between
Boroson\markcite{bor81} (1981) and Kent\markcite{ken85} (1985), and
the same analysis has been applied to them.  The percentage
differences between the fitting parameters from these works and the
present study have been computed and tabulated in Table 5.  In this
table, we list for each galaxy the percentage difference between the
parameters in the referenced work and this study, and also between the
two reference works; the last column identifies the reference work.
The differences were computed in the sense reference work minus this
study, and Boroson minus Kent.  These data are also presented in
Figure 5, which shows the fitting parameters from
Boroson\markcite{bor81} (1981) and Kent\markcite{ken85} (1985) plotted
against the values obtained in the present study.

In general, the fits from the various studies do not agree very well,
although the disk fits are typically more similar than the bulge fits.
The mean of the absolute values of the percentage differences are
105\% for $I_e$, 83\% for $r_e$, 61\% for $I_0$, and 26\% for $r_0$,
with 17 objects for all parameters.  Our bulge fits agree somewhat
better with those of Kent\markcite{ken85} (1985), while our disk
central surface brightnesses are closer to Boroson's\markcite{bor81}
(1981) results -- this reflects the conversion of Boroson's suface
brightnesses to the V-band. Interestingly, our values of $r_0$ agree
equally well, on average, with both reference works, leading us to
believe that a 25\% scatter in the disk scale length is to be expected
under the circumstances of this comparison (differing bandpasses, .

The bulge fit differences are dominated by the effects of the profile
acquisition procedures used in the different studies: this work used a
wedge-shaped major axis cut, while Boroson\markcite{bor81} (1981) and
Kent\markcite{ken85} (1985) used two variations on azimuthal
averaging.  In particular, it has been pointed out by
Boroson\markcite{bor81} (1981) that azimuthally averaged profiles will
be systematically too bright in the bulge-dominated regions due to
sampling the rounder bulge at a smaller galactocentric radius than the
disk, at a given position in the image.  Inspection of Table 5 reveals
that every instance of a large departure in $I_e$ shows a large
departure in $r_e$ of the {\em opposite }sign.  That our bulge
parameters agree better with Kent\markcite{ken85} (1985) than with
Boroson\markcite{bor81} (1981) probably reflects the use of fixed
ellipticities and position angles by Boroson, while Kent allowed those
quantities to vary with radius.  As a result, the mean surface
brightness around ellipses in bulge-dominated regions are more
representative of the major axis with Kent's profile acquisition
procedure than with Boroson's, hence the slightly better agreement
with our major axis cuts. 

\section{Fitting Characteristics}

The basic result of the profile decomposition process is that 620 of
the 659 profiles were successfully fitted with the chosen fitting
functions. In this section, we will discuss some of the
characteristics of the profile decomposition and the errors as they
appear with this sample of galaxies.  Our intent is to provide
information regarding the characteristics of the fitting; we will
present our analysis regarding the structure of disk galaxies
elsewhere.

\subsection{Fitting Parameters and Morphology}

All galaxies included in this sample have lenticular or spiral RC3
Hubble types; however, 61 of the fits have been made with no disk
component, indicating a lack of any appreciable exponential component
to their brightness profiles.  Visual inspection of these galaxies in
the PANBG shows many of them to have what appear to be disks even
though the profile shows none (e.g., NGC~1784, NGC~1961, NGC~3370);
others are possibly misclassified elliptical galaxies (e.g., NGC~2655,
NGC~3998, NGC~5485).  The galaxies with no disk component span a range
in T-Type from -3 to 9 --- the entire range of T-Type included in this
sample.  Early type galaxies are somewhat more likely to be fitted
without a disk than are late types, but the trend is not very strong.
There is no significant difference in the fit quality (as measured by
the tabulated RMS deviation) between the galaxies without disk fits
and the sample as a whole.

A total of 113 galaxies were fitted without a bulge component, and
although these galaxies range in T-Type from -2 through 9, they are
mostly late type galaxies.  This, of course, is consistent with the
basic behavior of the Hubble classification scheme where bulges are
less prominent in later types.  Again, the basic fitting quality is
the same for galaxies without bulge fits as with the sample in
general.

An investigation of the variation of the median fractional errors with
T-Type is summarized in Table 6.  This table gives for each range of
T-Type the median fractional error of each fitting parameter, as well
as the number of galaxies included in each median determination.  The
errors in $I_e$, $r_e$, and $I_0$ show some slight trends which are in
the expected senses, but which are small enough to be of questionable
significance.  The two bulge parameters seem to have somewhat larger
median errors for later types, as would be expected as the bulge
contribution to the light distribution decreases.  We should also note
the work of Andredakis and Sanders\markcite{and94} (1994), who show
that the inner regions of late-type spirals are perhaps better
represented by an exponential light distribution than a deVaucouleurs'
law.  Similarly, the disk central intensity error has larger values at
earlier types, when the disk contribution is generally lower.

We also looked at the possibility of a fitting quality dependence on
the presence or lack of a bar, and we have found nothing significant.
The median RMS residual of the fits on non-barred galaxies (Hubble
type in the RC3 contains an "A" explicitly) is about 0.14 magnitudes,
for barred galaxies ("B") it is 0.15 magnitudes, for mixed types ("X")
0.16 magnitudes, and for objects with no bar classification given in
the RC3 we find a value of 0.14 magnitudes.  All of these values are
sensibly the same as as the median value of 0.15 magnitudes found for
the full sample, and there is clearly no trend apparent.  Table 7
lists the median fractional error of the individual fitting parameters
for each bar class; there are no significant trends in these results.
We conclude that bars have no discernable effect on the quality of the
fits.

Objects which have been fitted with an inner truncation make up about
25\% (156/620) of the sample.  Remembering that inner truncations were
included only if they improved the {\em weighted} RMS by at least
10\%, this serves to justify our initial decision to use that function
with this large data set. Some of these inner-truncations are probably
caused by arms or rings at large galactocentric radii (e.g., NGC~2859,
NGC~3368, and NGC~5701) -- the arm/ring is bright relative to the
local disk and thus mimics an inner-truncation.  In these cases, the
arm/ring is typically faint (peaks near $24~V-mag~arcsec^{-2}$), and
has a short disk scale length.  These galaxies are also generally
classified as having an outer ring.  These objects are identified in
Table 1 with a "C" in the comment column.  A more-detailed analysis of
the presence of an inner-truncation is left for a later study.

\subsection{Fitting Parameters and Inclination}

An analysis of the unweighted RMS residuals shows little or no trend
in the mean value with inclination.  We divided the sample into three
inclination ranges based solely on the $R_{25}$ value listed in the
RC3, assuming that this isophote corresponds to a flat, circular disk:
$i \leq 30\arcdeg$ ($R_{25} \leq 0.0625$), $30\arcdeg < i \leq
60\arcdeg$ ($0.0625 < R_{25} \leq 0.301$), and $i > 60\arcdeg$
($R_{25} > 0.301$).  Note that no T-Type dependence was included in
this inclination estimate.  The resulting mean RMS residuals for the
low inclination, medium inclination, and high inclination samples are
0.16, 0.19, and 0.21 magnitudes, respectively.  These three values are
all much less than one standard deviation from each other, so the
trend is statistically meaningless. The median RMS residuals for each
inclination sample are 0.15 magnitudes, 0.14 magnitudes, and 0.17
magnitudes for the low, medium, and high inclinations, respectively.
We conclude that the sample has no significant inclination dependence
on the unweighted RMS residuals of the fits.

The quality of the individual fitting components is investigated by
computing the median fractional errors within each inclination range;
the mean fractional errors are not useful because of the outliers
discussed previously.  Table 8 provides the results of this
investigation, which shows that there are no indications of any
inclination dependencies.  In Table 8, the first column lists the
parameters, and the next three columns give the median relative errors
in each inclination range.

Finally, we check if the rate of occurrence of the inner-truncation
has any inclination dependence by computing the rate in each
inclination range.  The rates are $28\% \pm 6.5\%$ for the low
inclination group, $28\% \pm 3.3\%$ for the medium inclinations, and
$19\% \pm 3.3\%$ for the high inclination sample.  Thus, there is no
trend significant at the $2\sigma$ level.

\section{Conclusions}

We have presented one of the largest, if not the largest, collections
of spiral and lenticular galaxy brightness profile bulge-disk
decompositions yet completed.  Of the 659 brightness profiles in our
sample, 620 were fit with the deVaucouleurs law plus inner-truncated
exponential disk function, while the remaining 39 profiles could not
be so fit.  The general quality of the fits is quite high, with about
50\% of the fits having an unweighted RMS deviation from the data
(including real structures) of less than 0.15 magnitudes and more than
90\% have unweighted RMS residuals of less than 0.35 magnitudes.  We
find no systematic trends in the fitting quality with either galaxy
morphology or inclination.  Comparison of our fits with those of
Boroson\markcite{bor81} (1981) and Kent\markcite{ken85} (1985) show
discrepancies attributable to a number of observational and data
reduction factors.

Probably the most interesting result from this process of fitting is
simply that we achieved a "success" rate of 620/659 (94\%) in our
fits, compared to success rates of 75/94 (80\%) for
Kent\markcite{ken85} (1985; spiral and S0 galaxies only) and 16/26
(62\%) for Boroson\markcite{bor81} (1981).  While these success rates
are statistically similar, we wonder if the small number statistics
are the only differences.  An analysis of the 19 non-fittable disk
galaxy profiles from Kent\markcite{ken85} (1985) shows that we found
fits for all ten of those galaxies which were also in our sample. For
these 10 objects, the mean inclination is 41$\arcdeg \pm$ 18$\arcdeg$,
the median is 42$\arcdeg$, and only one galaxy (NGC~5566) has an
inclination greater than 60$\arcdeg$.  Thus, Kent's non-fittable
galaxies do not tend to be high inclination objects.  Furthermore, the
RMS deviations of our fits for these same galaxies are generally
small, with a median value of 0.15 magnitudes, the same as for our
full sample.  The princpal difference seems to be that we included the
inner-truncated disk (ITD) factor in our fits: seven of these 10
galaxies have ITD's in our fits, often with large $r_{h}/r_{0}$
ratios.  It is also possible that Kent's\markcite{ken85} simultaneous
fitting of the minor axis profiles made his results more sensitive to
deviations from the standard fitting functions.  A similar analysis of
the 10 non-fittable profiles in Boroson's\markcite{bor81} (1981)
sample shows us having fits for the nine which are included in our
sample.  These nine galaxies are also of relatively low inclination
(the largest is about 58$\arcdeg$), and have small RMS deviations in
our fits (median value of 0.13 magnitudes).  However, our fitted
parameters show three objects with bulge only, three with a bulge plus
exponential disk, and three with a bulge plus ITD; the case for the
inclusion of the ITD is not as strong with this set of profiles.  We
conclude, however, that the inclusion of the ITD term in our fitting
functions has allowed us to fit an additional 10\% to 15\% more
galaxies than we would have fit without the inner-truncation term.

It is also interesting that about 25\% of the profiles in our sample
are fitted better by an inner-truncated disk function than with a
plain exponential.  Some of these fits are certainly due to the ITD
being fitted to outer rings (as indicated by a "C" in the last column
of Table 1), and others may be marginal improvements (remember the
requirement for a 10\% improvement of the {\em weighted} RMS to
include the inner-truncation term), but clearly a significant fraction
of the profiles support the physical reality of the inner-truncation
in the light distribution.  A quantification of the strength of the
inner truncation and the search for the origin of this feature is the
subject of a future paper.

Acknowledgements: One of the authors (WEB) has been supported by STScI
under contract NAS5-26555 for this work.  The authors would like to
thank M. Hamabe for making the PANBG brightness profiles available to
us for this project.  We also wish to thank the anonymous referee for
some useful suggestions.  Part of the data analysis for this paper
used STSDAS, which was developed by the Space Telescope Science
Institute under U.S. Government contract NAS5-26555.

\clearpage

\appendix
\section{Effects of Seeing on the Fitting Results}

The fitting procedure described in section 3.1 makes no allowance for
the effects of seeing other than to start the fitting at a radius
larger than $3\arcsec$ to avoid the most-affected portion of the
brightness profile.  We describe in this Appendix a set of experiments
which were used to derive estimates of the correction factors for the
fitted parameters to measure more accurately the true parameters of
the bulge light distributions.

There are four factors in the profile acquisition process which can
influence the fitting results, all of which occur at distinct stages
of the process and which can be assumed to be separable.  Seeing
smears the galaxy light as it travels through the atmosphere, while
Poisson noise occurs during the photographic exposure physics and
chemistry.  Pixellation broadens sharp features during the
digitization of the plate, as well as adding some additional Poisson
noise, and smearing by the aperture photometry happens as a result of
the profile acquisition from the digitized data.  Our experiments were
designed to investigate only those processes which broaden sharp
features of the galaxy light distribution: seeing, pixellation, and
aperture photometry.

The PANBG provides seeing estimates for all plates in their Table 4.1;
we assume that "seeing" in this case refers to the FWHM of stellar
profiles.  The tabulated seeing in the PANBG ranges from $1\arcsec$ to
$7\arcsec$ and is included in our Table 1.

We attempted to replicate the profile definition process as accurately
as possible utilizing computer-generated images.  These images consist
of a perfectly circular deVaucouleur's light distribution with input
values of $I_{e}=1000$ in arbitrary intensity units and $r_{e}= 0.5,
1, 2, 4, 8, 16,$ and $32\arcsec$.  Poisson noise was included in the
images.  No disk component was included in these images because the
most pronounced effects of seeing will be on the bulge component due
to its very steep slope at small radii.  We generated 512x512 pixel
images with these characteristics, assuming a pixel scale of
3~pixels/$\arcsec$ (to simulate the photographic resolution), then
smeared them with Gaussians of FWHM=1, 2, 3, 4, 5, 6, and $7\arcsec$
to simulate the effect of the atmosphere during a long exposure.  The
resulting images were then block averaged with a 3x3~pixel ($1\arcsec$
square) block to replicate the plate scanning utilized for most of the
galaxies in the PANBG.  Finally, the brightness profiles from the
simulated images were extracted using a set of variable diameter
circular apertures along a single radius of the "galaxies." These
profiles were then fitted in the same fashion as the real profiles in
order to estimate the functional parameters, using a fixed fitting
range of $3\arcsec$ to $48\arcsec$.  A second set of images was
generated using the same input parameters, but which were not smeared
by the Gaussians; these images were used to determine the correction
factors due to the pixellation and profile acquisition aperture
photometry prior to estimating the effects of smearing with a
Gaussian.

The results of this procedure are presented in Table 9, which
includes for each simulation the seeing size in arcseconds, the input
$I_{e}$ and $r_{e}$, the fitted $I_{e}$ and $r_{e}$, and the ratio of
each fitted parameter to its input value, all based on the
fully-degraded brightness profiles.  The $I_{e}$ are all in arbitrary
intensity units, and the $r_{e}$ are all in simulated arcseconds.  Figure A1
shows the relationship between the fitted:input $I_{e}$ as a function
of the fitted:input $r_{e}$; the two quantities are highly
anti-correlated, suggesting that the fitted parameters are not truly
independent of one another.

As can be seen from the table, there are some cases where the fitted
parameters are very different from the input values: the worst cases,
as would be expected, are where the seeing disk is large compared to
the input $r_{e}$.  For $r_{e}$, the percent errors range from less
than 1\% to almost 40\% (seeing=$7\arcsec$, $r_{e}=1\arcsec$), while
for $I_{e}$ the range is from almost 0\% (seeing=$4\arcsec$,
$r_{e}=32\arcsec$) to more than 1000\% (seeing=$7\arcsec$,
$r_{e}=1\arcsec$).  We have every reason to suspect that similar
fitting errors exist in the PANBG fits, and a means of correcting
these errors is highly desirable.

The correction procedure has been separated into two stages: first
correct the fitted parameters for the combined effects of pixellation
and aperture photometry, then correct these modified parameters for
the effects of the Gaussian smearing.  The pixellation/aperture
photometry correction for the effective radius, based on the
simulations without Gaussian smearing, has been found to be
well-represented by a power law of the form

\begin{equation}
log(r'_{\rm e}/r_{\rm e}) = 0.005 log(b/r_{\rm e}) - 0.018
\end{equation}

\noindent 
or

\begin{equation}
r'_{\rm e}=0.959 r_{\rm e} (b/r_{\rm e})^{0.005}
\end{equation}

\noindent 
where $r_{\rm e}$ is the fitted effective radius, $b$ is the
digitization aperture size ($1\arcsec$), and $r'_{\rm e}$ is the
effective radius corrected for the effects of pixellation and
aperture photometry; the units of all quantities are in arcsec.  This
function has a correlation coefficient of 0.989, and a maximum
percentage error of 0.2\% within the parameter space studied.

The pixellation/aperture correction function for the effective
intensity is similarly found to be

\begin{equation}
I'_{\rm e}=[0.068 log(b/r_{\rm e}) + 1.246]I_{\rm e}
\end{equation}

\noindent 
where $I_{\rm e}$ is the fitted effective intensity, $r_{\rm e}$ is
the fitted effective radius, $b$ is the digitization aperture size,
and $I'_{\rm e}$ is the effective intensity corrected for pixellation
and aperture photometry.  This function has a correlation coefficient
of 0.995 and a maximum percentage error of 0.6\% within the parameter
space studied.  Figure A2 shows the fits of the pixellation/aperture
photometry correction functions for both $r_{\rm e}$ and $I_{\rm e}$.

The seeing correction is applied by the linear interpolation of the
galaxy parameters on the model parameters for the same value of
seeing.  In Table 10 we list the model data used in the interpolation;
column 1 lists the seeing in arcsec ($S$), column 2 the ratio of the
seeing to the pixellation-corrected effective radius ($S/r'_{\rm e}$),
column 3 the ratio of the input effective intensity to the
pixellation-corrected effective intensity ($I^{\rm 0}_{\rm e}/I'_{\rm
e}$), and column 4 the ratio of the input effective radius to the
pixellation-corrected effective radius ($r^{\rm 0}_{\rm e}/r'_{\rm
e}$).  The interpolation procedure is performed by computing the
$S/r'_{\rm e}$ ratio for a fit, and interpolating the $I^{\rm 0}_{\rm
e}/I'_{\rm e}$ and $r^{\rm 0}_{\rm e}/r'_{\rm e}$ ratios for the
tabulated seeing disk size to derive the fully-corrected $I^{\rm
0}_{\rm e}$ and $r^{\rm 0}_{\rm e}$ from the previously-computed
values of $I'_{\rm e}$ and $r'_{\rm e}$ (eqs. A2 and A3).

We include in Table 1 the corrected bulge parameters resulting from
the described correction procedure.  Corrections for objects whose
fitted parameters lie outside of the interpolation parameter space are
not provided in Table 1.

\clearpage

\figcaption[wbaggett_fig1a.ps,wbaggett_fig1b.ps]{Brightness profile
fits to the NGC~3145 data. Both plots show the bulge component, the
disk component, and the sum of the two. The horizontal line at 25.3
$V-mag~arcsec^{-2}$ indicates the range of radius included in the fit.
{\em Left:} Fit without a truncation term in the disk component; the
RMS deviation of this fit is 0.28~mag.  {\em Right:} Fit including the
truncation term in the disk component; the RMS deviation of this fit
is 0.12~mag.}

\figcaption[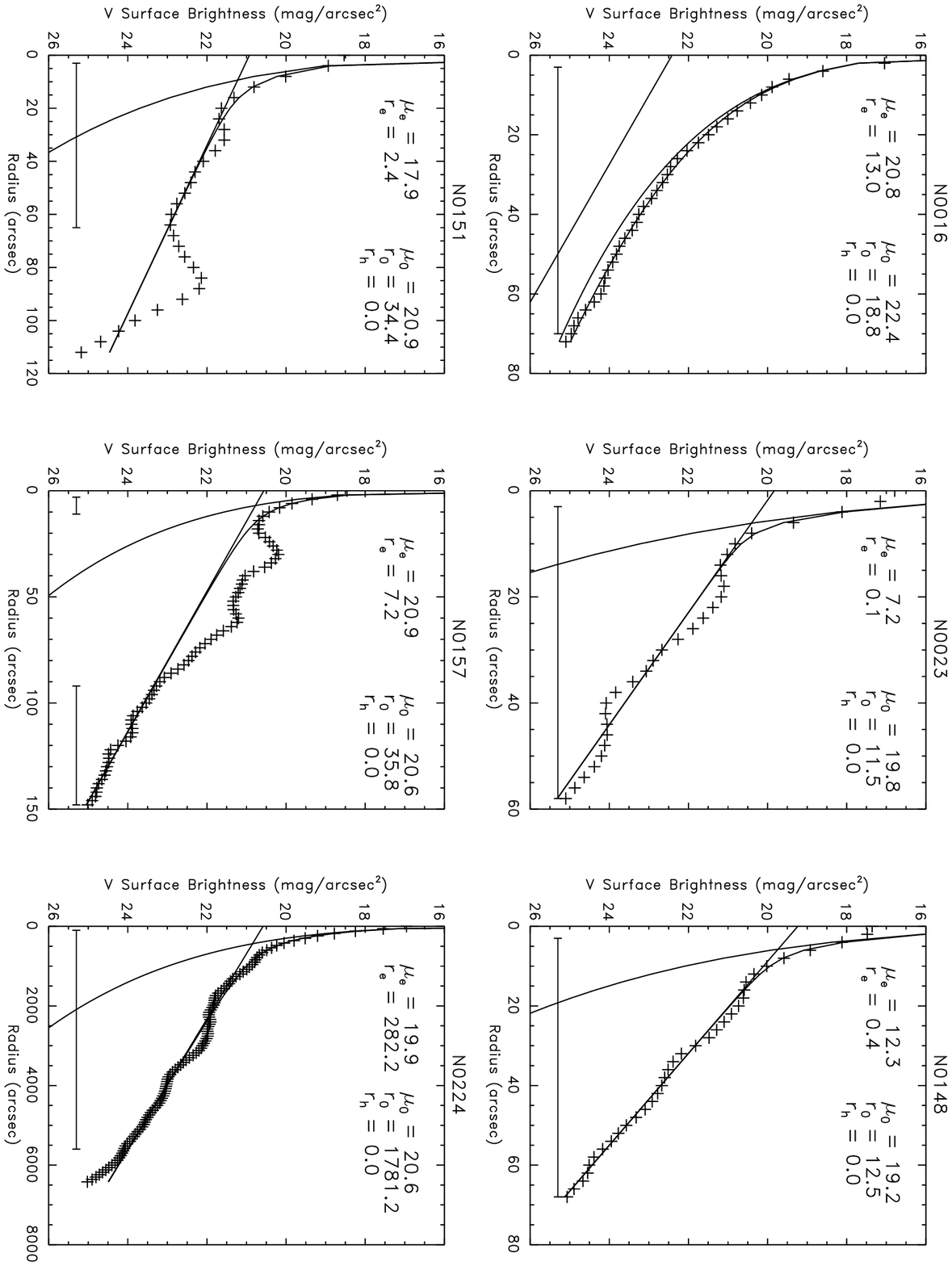]{Brightness profile fits to the galaxies
presented in Table 1.  Each plot shows the observed profile (crosses),
the bulge and disk components and their sum (solid lines), and the
range of radius included in the fit (horizontal line(s) at bottom).
Also shown are the fitted parameters, where the surface brightness
parameters are in units of $V-mag~arcsec^{-2}$ and the size parameters
are in arcseconds.}

\figcaption[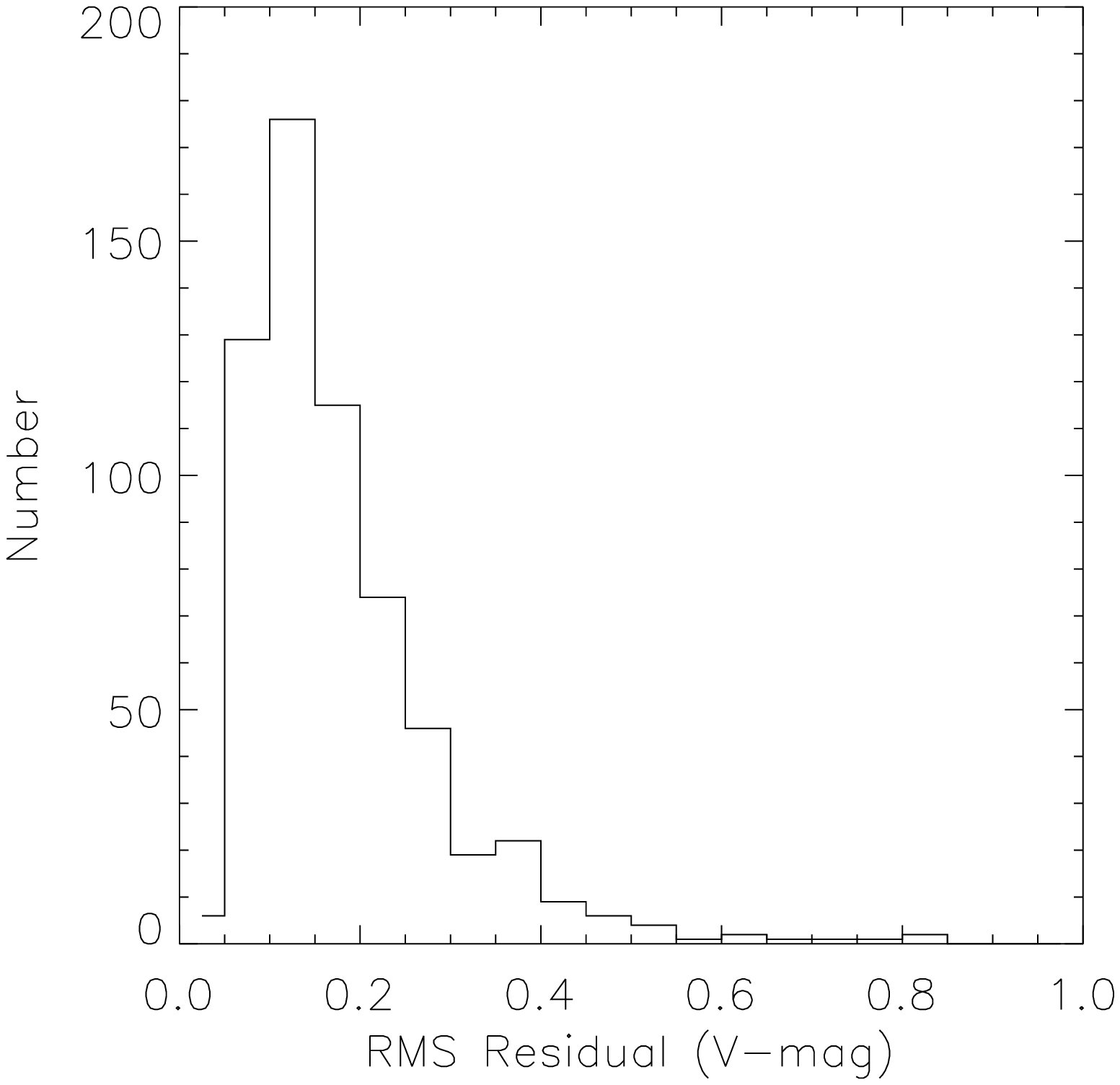]{Distribution of RMS deviations of the
fits from the data.  The RMS deviations were computed for all data
points in each profile between 3~arcsec and the first point at which
the surface brightness drops below $25~V-mag~arcsec^{-2}$.  The median
value of the distribution is at 0.15~mag. There are six objects which
have RMS deviations greater than 1.0~mag which are not included in the
figure.}

\figcaption[wbaggett_fig4.ps]{Histograms of the fractional errors.
The distributions of the estimated fractional errors of each fitted
parameter are shown to illustrate the problem with outliers: a)
$I_{e}$, b) $r_{e}$, c) $I_{0}$, d) $r_{0}$, and e) $r_{h}$.  Note the
smaller bin size for $r_{0}$ and $r_{h}$.}

\figcaption[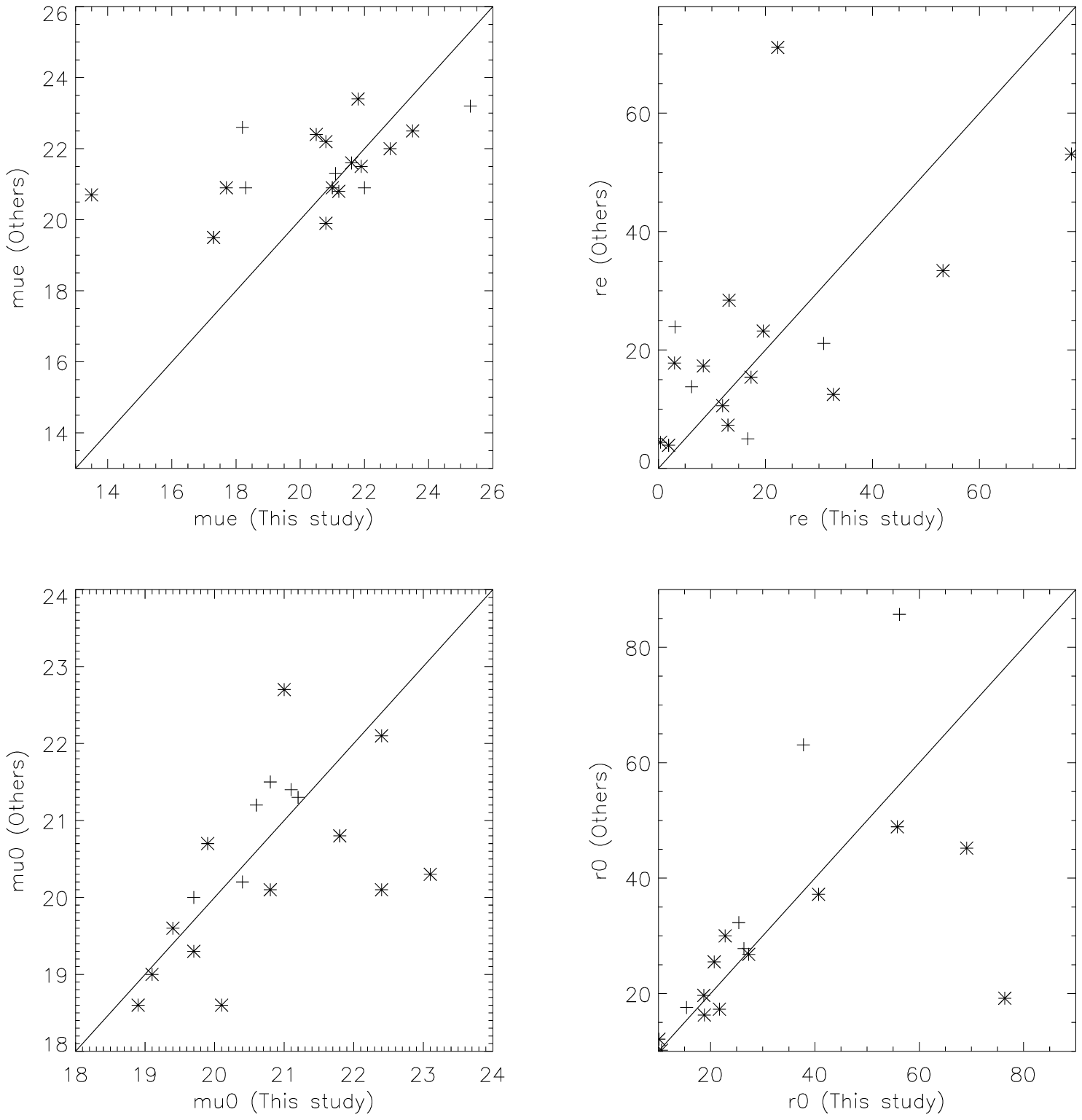]{Comparison of fitted parameters.  The
fit parameters for the galaxies in common with Boroson (1981) (plus
signs) and Kent (1985) (asterisks) are shown along with the 45-degree
line.  The surface brightness parameters are expressed in units of
$mag~arcsec^{-2}$ and the length parameters are given in $arcsec$.
There is a reasonable correlation for all of the parameters, although
the disk parameters have tighter correlations than those for the
bulge.}

\figcaption[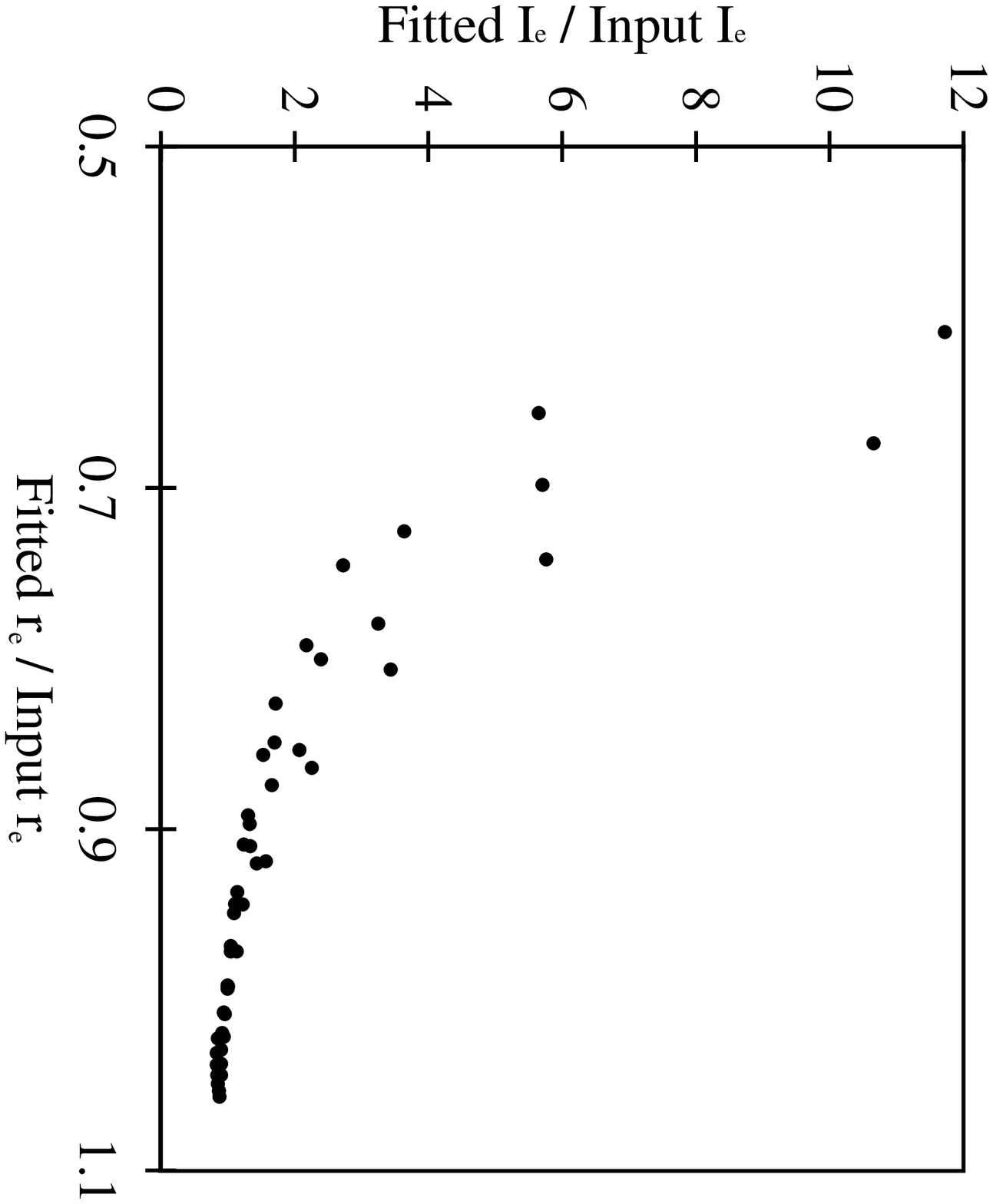]{Relationship between $I_{\rm e}$
fitting accuracy and $r_{\rm e}$ fitting accuracy.  The apparent
anticorrelation suggests some dependence of the fitting parameters on
each other.}

\figcaption[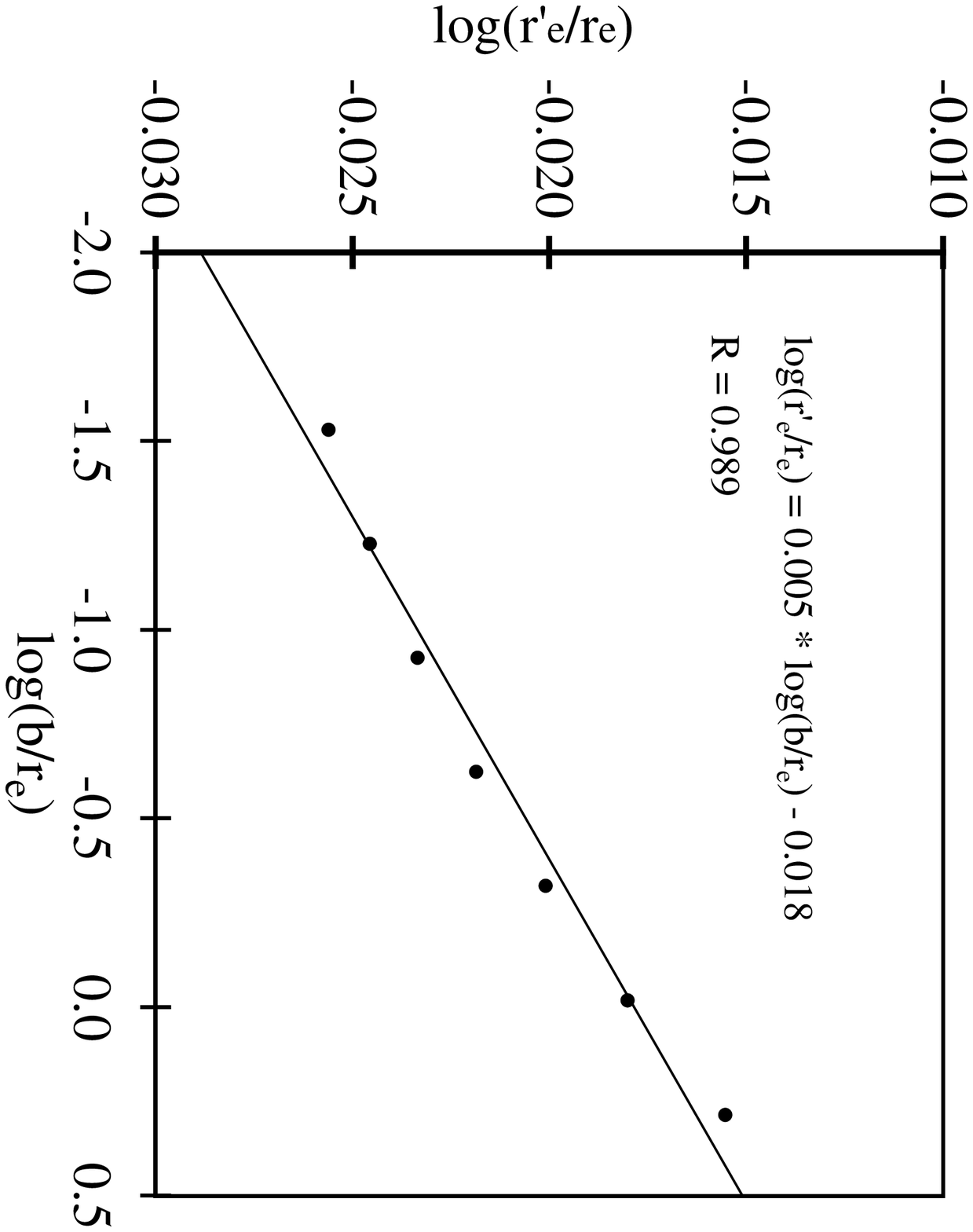,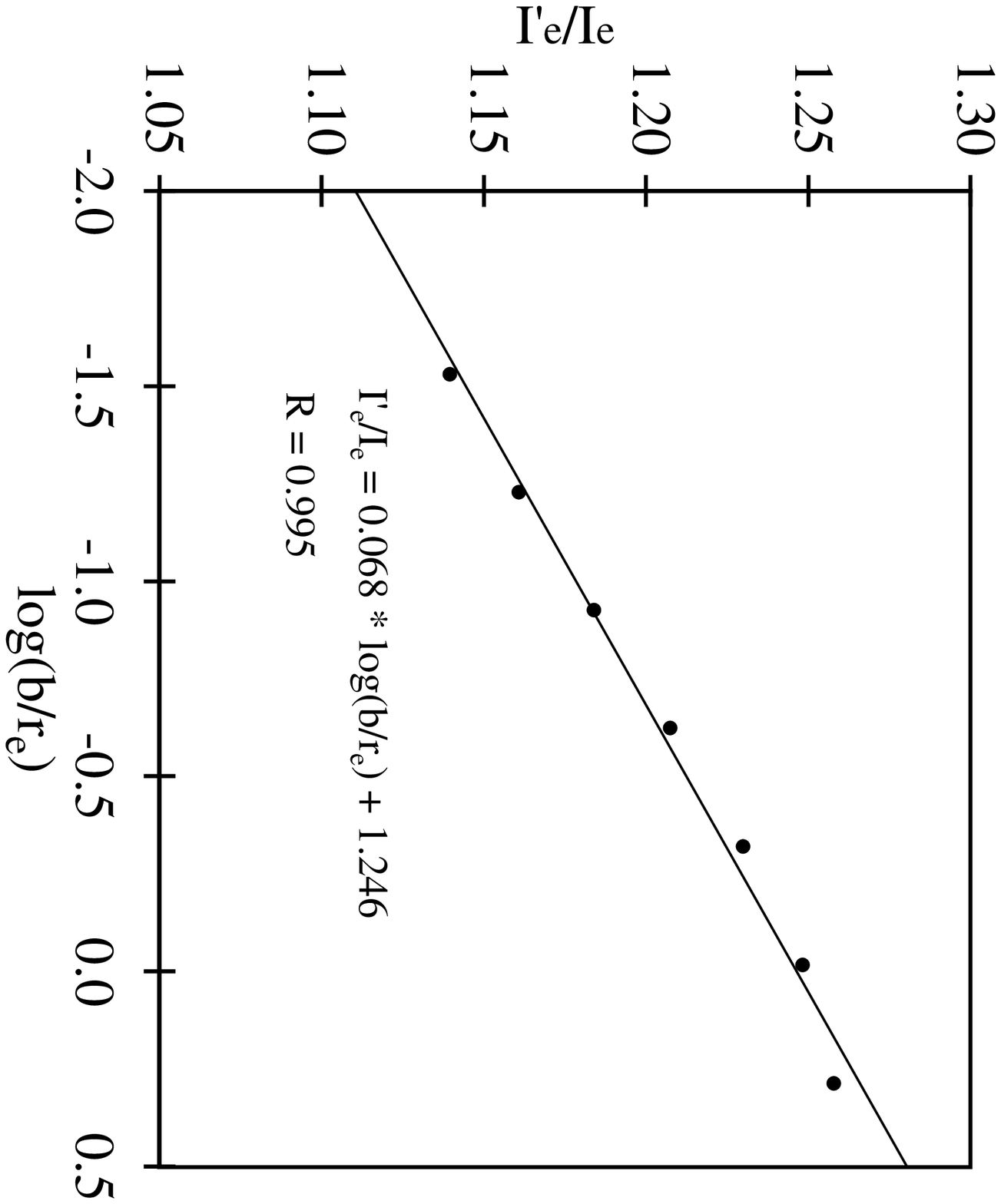]{Pixellation/aperture
photometry correction functions. a) The data and fit for the
correction function for $r_{\rm e}$ is shown in a log-log plot.  There
is clearly a systematic residual function, but it is of insignificant
amplitude.  b) The data and fit for the $I_{\rm e}$ correction function
is shown; again there is an insignificant systematic residual,
particularly at large values of $b/r_{\rm e}$.}

%
 
\end{document}